# Reconciliation of experiments and theory on transport properties of iron and the geodynamo

Youjun Zhang[1,2*], Mingqiang Hou[2,3*], Guangtao Liu[2], Chengwei Zhang[2], Vitali B. Prakapenka[4], Eran Greenberg[4], Yingwei Fei[5], R. E. Cohen[5,6★] & Jung-Fu Lin[7★]

[1]Institute of Atomic and Molecular Physics, Sichuan University, Chengdu 610065, China.
[2]Center for High Pressure Science and Technology Advanced Research (HPSTAR), Shanghai 201900, China.
[3]The Advanced Light Source, Lawrence Berkeley National Laboratory, Berkeley, CA 94720, USA.
[4]Center for Advanced Radiation Sources, University of Chicago, Illinois, USA.
[5]Extreme Materials Initiative, Geophysical Laboratory, Carnegie Institution for Science, Washington, DC 20015-1305, USA.
[6]Department of Earth and Environmental Sciences, LMU Munich, Theresienstrasse 41, 80333 Munich, Germany.
[7]Department of Geological Sciences, Jackson School of Geosciences, The University of Texas at Austin, Austin, TX 78712, USA.

★E-mail address: afu@jsg.utexas.edu (J.F.L.), rcohen@carnegiescience.edu (R.E.C.)
*These authors contributed equally to this work.

**Abstract**

The amount of heat transport from the core, which constrains the dynamics and thermal evolution of the region, depends on the transport properties of iron[1,2]. Ohta et al. (2016) and Konôpková et al. (2016) measured electrical resistivity[3] and thermal conductivity[4] of iron, respectively, in laser-heated diamond anvil cells (DACs) at relevant Earth's core pressure-temperature (*P-T*) conditions, and obtained dramatically contradictory results[5,6]. Here we measure the electrical resistivity of *hcp*-iron up to ~170 GPa and ~3,000 K using a four-probe van der Pauw method coupled with homogeneous flat-top laser-heating in a DAC. We also compute its electrical and thermal conductivity by first-principles methods including electron-phonon[7-9] and electron-electron scattering[10,11]. We find that the measured resistivity of *hcp*-iron increases almost linearly with increasing temperature, and is consistent with current first-principles computations[11]. The proportionality coefficient between resistivity and thermal conductivity (the Lorenz number) in *hcp*-iron differs from the ideal value ($2.44\times10^{-8}$ W$\Omega$K$^{-2}$), so a non-ideal Lorenz number of ~$(2.0 - 2.1)\times10^{-8}$ W$\Omega$K$^{-2}$ is used to convert the experimental resistivity to the thermal conductivity of *hcp*-Fe at high P-T. The results constrain the resistivity and thermal conductivity of *hcp*-iron to ~80±5 μΩ·cm and ~100±10 W/m·K, respectively, at conditions near core-mantle boundary. Our results indicate an adiabatic heat flow of ~10±1 TW through the core-mantle boundary for a liquid Fe alloy outer core, supporting a present-day geodynamo driven by thermal convection through the core's secular cooling and by compositional convection through the latent heat and gravitational energy during the inner core's solidification.

**Main text**

Earth's core works like a heat engine through heat transfer from the cooling and freezing of the liquid iron core, which powers the present-day geodynamo, mantle convection, and plate tectonics[12,13]. Paleomagnetic records indicate that the geodynamo has been active for at least 3.4 billion years[14]. The geodynamo was long believed to be driven by primordial heat in the earth being transported by thermal convection through the liquid outer core. Heat flow of 3 − 4 TW across the core-mantle boundary (CMB) may suffice to sustain the geodynamo from the early history of Earth's core, indicating an old inner core of ~3.5 Gyr[15]. However, first-principles calculations by de Koker et al. (2012) and Pozzo et al. (2012) claimed a much higher adiabatic heat flow (15 − 20 TW) at the CMB, and estimated thermal conductivity $\kappa$, of 150 Wm$^{-1}$K$^{-1}$ in pure Fe[7,8]. Their conductivity value is three to five times higher than geophysical estimates and shock-wave experiments[15,16,17,18]. This discrepancy ignited a debate on the geodynamo, Earth's energy budget, and thermal evolution of the core. Present scenarios include convection driven by chemical differentiation rather than thermal convection[19], which requires a different process for each planetary dynamo, as opposed to thermal convection, which acted as a universal concept for driving planetary dynamos. Nevertheless, thermal evolution models show that even these higher conductivities can be consistent with a thermal driven dynamo when radiogenic heating of the core is included[20]. The conductivity values of iron in the core also influence our assessment of the age of the inner core formation: the higher the electrical and thermal conductivity, the faster the Earth's core cools, and the younger the inner core[12].

Several experimental studies have been conducted at high *P-T* conditions, but measurements of the transport properties of Fe using different methods came to dramatically contradictory results[2-4,21,22]. The current estimate of thermal conductivity of iron under core-mantle boundary (CMB) conditions varies by a factor of ~6 in laser-heated DAC experiments[23], which translates into the CMB heat flux of 4–20 TW. Ohta et al. (2016) and Suehiro et al. (2019) measured the electrical resistivity (the reciprocal of electrical conductivity) of an Fe wire in a laser-heated and internally resistive-heated DAC, respectively, and inferred strong resistivity saturation at high temperatures, indicating a very high thermal conductivity of $226^{+71}_{-31}$ Wm$^{-1}$K$^{-1}$ in *hcp*-Fe at the topmost outer core if the Wiedemann-Franz relation between resistivity and thermal conductivity is applied with an ideal Lorenz number ($L_0$ = 2.44 × 10$^{-8}$ WΩK$^{-2}$). Konôpková et al. (2016)

measured the thermal conductivity of Fe by observing a heat pulse across a hot dense Fe foil using a dynamically laser-heated DAC. Contrary to the inferences from the resistivity measurements, they obtained a low thermal conductivity of 33 ± 7 Wm$^{-1}$K$^{-1}$ near the CMB, which agrees with the early estimation[16]. Recent thermal diffusivity measurements in *hcp*-Fe by pulse light heating thermoreflectance in a laser-heated DAC seems consistent with the low thermal conductivity at ~50 GPa and ~1,400 K[24]. New theoretical calculations show that both electron-phonon and electron-electron scatterings contribute to the thermal conductivity of iron, and give a value of 97 ± 10 W m$^{-1}$ K$^{-1}$ for pure *hcp*-Fe at the CMB conditions[11]. Therefore, the electronic correlations in the transport property calculation of Fe should be considered[25]. This value is between the estimates from the experimental thermal conductivity and electrical resistivity measurements.

The discrepancy[5,6] between the thermal conductivity and resistivity measurements may arise from temperature gradients in the laser-heated samples, sample deformation and textures, and probe geometry issues. In Ref. 3, for example, the broad laser-spot (~30 μm) with a Gaussian beam shape in the laser-heated DAC may introduce a temperature gradient in the sample's heating region. In addition, the two-probe sample geometry they used (pseudo four-probe) is insufficient to reliably measure the resistance of a micro sheet sample[26], so uncertainties in the derived resistance (refs. 3 and 22) are expected to be significant. Furthermore, the ideal Lorenz number in the Wiedemann-Franz relation ($\kappa = L_0 T \sigma$) was used to derive the thermal conductivity at CMB conditions, but this may be inappropriate[7,8]. In terms of theory, most previous computations considered only the contribution of electron-phonon (*e-ph*) scattering to the electrical resistivity[7,8], but recent studies indicate that electron-electron (*e-e*) scattering play an important role as well, especially at high temperatures[10,11,27]. Thus, further improved and integrated experimental measurements and theoretical studies on the transport properties of Fe at high *P-T* are needed to understand the actual conductivity of the core, which in turn helps us to accurately assess Earth's heat budget, the geodynamo, the age of the inner core, and the thermal evolution of the Earth's core and lowermost mantle. Uncertain thermal conductivity of the core may result in a big difference in our understanding of the thermal history of the core and its geodynamo[28,29].

We used a modified four-probe van der Pauw method[30] to reliably measure the electrical resistivity of *hcp*-Fe at high *P-T* in a double-side laser heated DAC with two flat-top laser beams[31] (See Methods). An iron sample (~2 μm thick) was shaped into a uniform Greek cross sheet with a circular region of approximately 6 μm at the central cross area (cloverleaf) using a Focus Ion

Beam (FIB) (Fig. 1a). The cross sheet has four arms with a length of ~70 μm, which act as internal electrode wires (Fig. 1b). The arms were connected to four external Pt leads. The heart area of the cross sheet with a small disc with ~6 μm in diameter matches well with a finely focused laser-heating spot (~10 μm) as shown in Fig. 1c. Our samples have a smaller width than the laser spot, ensuring homogeneous flat-top heating with uniform temperature distributions on both sides of the sample. The Greek cross sheet with a diameter varied from ~6 to 100's μm has been proved to be a valid van der Pauw test structure by theory and experiment[26,30]. The sample assembly also avoids contact resistance and contamination from possible alloying between the Pt leads and the Fe sample.

We first measured the electrical resistivity of *hcp*-Fe at high pressures up to ~140 GPa at room temperature. The pressures and volumes were determined by the thermal equation of state[32] of *hcp*-Fe from *in situ* synchrotron X-ray diffraction at high *P-T* (example patterns are shown in Extended Data Fig. 1) during the electrical resistivity measurements at GSECARS, Advanced Photon Source. The resistivity in *hcp*-Fe decreases with increasing pressure (Extended Data Fig. 2), consistent with previous studies[2,21]. We then measured resistivity at ~105 GPa with increasing temperature up to ~3,000 K (Fig. 2); the resistivity increases with increasing temperature almost linearly, with an intercept of 4.8 μΩ cm at room temperature. The temperature dependence of our data is very different from the result of previous experiments[3], which showed significant reduction of resistivity with temperature (Fig. 2). Our measured resistivity is around 1.6 times higher than Ref. 3 up to 3,000 K at ~105 GPa. This difference is likely due to the temperature gradient and sample geometry in the earlier experiments. We tested this at ~74 GPa by introducing an artificial temperature-gradient, in which we heated one corner of a relatively large sample of ~15 μm in diameter compared to the focused laser spot of ~10 μm (Extended Data Fig. 3a). When the laser spot size and position did not match well with the sample, we observed strong resistivity saturation similar to what was reported in Ref. 3 (Extended Data Fig. 3b). The details on the temperature gradient issue are discussed in the Supplementary Information.

We carried out additional high-temperature experiments using homogeneous laser-heated DACs up to ~3,000 K from ~82 GPa to ~165 GPa, where only *hcp*-Fe was observed (Extended Data Fig. 4). We analyzed the temperature distribution on a sample with ~6 μm width at ~2,380 K and ~142 GPa. It shows a very homogeneous temperature in a width of ~8 μm (Extended Data

Fig. 5), which matched well with the sample size. The measured resistivities at 82, 133, 142, and 165 GPa are shown in Extended Data Figs. 6a to 6d, respectively. We found that the resistivity for all these well-heated Fe samples in our experiments increased quasi-linearly with increasing temperatures up to ~3,000 K in experiments. The results could be fitted well with the Bloch-Grüneisen formula at the measured *P-T* range (solid blue curves) (See Methods and Extended Data Table 1). Of particular interest is the parameter "*n*", which mainly depends on the nature of phonon and electron scattering. It is less than 5 and decreases from 3.66 to 0.81 with pressure change from 82 to 165 GPa, indicating electron-electron interactions also contribute to the resistivity in *hcp*-Fe at high pressures and temperatures. Our results exhibit a similar trend in "*n*" with the previously fitted Bloch-Grüneisen formula from resistivity measurements in a muffle furnace DAC up to 450 K, indicating that the temperature dependence of resistivity becomes weaker with increasing pressure[3].

We compare our experimental data with computations of transport properties that include *e-ph* and *e-e* scattering based on first-principles lattice dynamics (FPLD) and density functional perturbation theory (DFPT)[11], and also with our new results using first-principles molecular dynamics (FPMD) (Supplementary Information and Extended Data Fig. 7). The *e-ph* contribution was calculated using the inelastic Boltzmann transport equation[33] and DFPT; and the *e-e* contribution was obtained using density functional theory and dynamical mean field theory (DFT + DMFT)[34,35]. Any saturation effects are included automatically and naturally in the FPMD (See Extended Data Figure 7 and discussion). Neither the present calculations, nor those in Ref. 11 use fitting to data to obtain saturation effects, but the previous work relied on a model of saturation based on mean-free path and assumed Matthiessen's rule, that *e-e* and *e-ph* scattering can be computed separately and added. Our new measurements are consistent with the computed resistivities by DFPT and DMFT (ref. 11) (solid red squares and dashed red line in Fig. 2). In addition, the calculation is also consistent with a model that assumes a Bloch-Grüneisen form fit to our data (Fig. 2). Using separately computed *e-ph* and *e-e* scattering with a model for saturation effects in Ref. 11 gives about 10% higher resistivity than that extrapolated from the Bloch-Grüneisen formula up to 3,500 K. Using first-principles molecular dynamics with DFT/DMFT we obtain new results that almost agree perfectly with our new experimental results for solid *hcp* iron in Fig. 2 (solid green star). It is encouraging that a very different theoretical technique based on FPMD and very different assumptions agrees well with FPLD results that include a model for resistivity saturation. The measured resistivity of *hcp*-Fe is also compared with the calculated resistivity only contributed by *e-ph* scattering (ref. 11, open red squares and dash-dotted red line in Fig. 2). One can see that the *e-e* contribution to the resistivity is less than 10% at ~1,500 K, but it reaches above ~20% to 3,000 K at ~105 GPa, indicating a non-negligible effect of *e-e* scattering on the resistivity of *hcp*-Fe by both the experiments and theories, especially at high temperatures. The effects on thermal conductivity are larger, as discussed in Ref. 11, indicating a non-ideal Lorenz number at high P-T.

To estimate transport properties for Earth's core, we extrapolated the resistivity of *hcp*-Fe up to 4,000 K above 133 GPa by the Bloch-Grüneisen formula (Fig. 3a and Extended Data Fig. 6). The electrical resistivity is found to be ~80 ± 5 μΩ·cm in *hcp*-iron near CMB conditions (~136 GPa and 4,000 K). Compared with the previous experimental data (ref. 3), our results show about 1.5 to 2 times higher resistivity at the relevant conditions of the outer core (Fig. 3a). Based on the computed electrical resistivity and thermal conductivity of *hcp*-Fe at the CMB conditions (ref. 11), the Lorenz number is estimated to be ~(2.0 − 2.1) × $10^{-8}$ WΩK$^{-2}$, which is 20% lower than the ideal value (Supplementary Information and Extended Data Fig. 8). Comparing the Lorenz number with that obtained by de Koker et al. (2012) (open diamonds in Fig. 3b) shows that our finding is ~10% lower than that calculated from molecular dynamics simulation without *e-e* scattering for Fe. We convert the experimental resistivity data for *hcp*-Fe at high *P-T* to the thermal conductivity using the computed Lorenz number in the Wiedemann-Franz law, and obtain a thermal conductivity significantly lower than the previous estimates by Ohta et al (2016) that used the ideal Lorenz number (2.44 × $10^{-8}$ WΩK$^{-2}$) (Fig. 3b). At about 136 GPa and 4,000 K, at near CMB conditions, we find that the thermal conductivity in *hcp*-Fe is around 100 ± 10 W/m/K. Our values are still somewhat higher than the results measured by Konôpková et al. (2016) through direct observations of heat pulse in hot dense *hcp*-Fe, and deviate more with their extrapolated values with increasing pressure. Thus, it seems that the previous electrical conductivity was too high, and the previous thermal conductivities were too low. Our theory and experiment results fall between the above two previous studies[3,4]. The effect of *hcp*-Fe textures in a DAC on the electrical anisotropy is estimated to be approximately 10%, whereas recent calculations show the electrical anisotropy between *c* and *a* axis to be approximately 26% at Earth's core P-T conditions[11]. A recent study reported approximately 30% anisotropy in modeled thermal conductivity of textured *hcp*-Fe samples at 20–45 GPa and 300 K[36], which are generally consistent with the first-principles estimates (ref. 11). A very large anisotropy of $k_c/k_a$ = 3–4 for *hcp*-Fe at relevant core *P-T* conditions was suggested by an extrapolation of the limited data to Earth's core, but the uncertainty in the extrapolation is too large to be credible for our understanding of the core geodynamo[36]. Therefore, the discrepancy between this study and previous experimental values cannot be simply explained by textures of *hcp*-Fe crystals (Supplementary Information).

The outer core is liquid so the effects of melting on the thermal conductivity should be considered. Previous high *P-T* experiments in a heated multi-anvil apparatus show a ~5 − 10% resistivity increase upon melting from *fcc*-Fe below ~10 GPa[37-40]. It is still difficult to directly measure the resistivity of liquid Fe in experiments at core conditions so that we have to estimate the effects of the melting of *hcp*-Fe. The recent computed resistivity and thermal conductivity considered scattering of electrons from both atomic motions (*e-ph*) and electrons (*e-e*) shows a 7 − 10% increase in resistivity or decrease in thermal conductivity upon melting from *hcp* structure at 4000 − 4500 K and ~145 GPa (ref. 11), which is slightly lower than previous calculations of ~15% change, which only considered *e-ph* contributions (ref. 9). Therefore, if we use a value of ~10% increase of electrical resistivity after melting, the thermal conductivity of liquid Fe would be around 90 ± 15 W/m/K at the relevant condition of the core-mantle boundary (solid red star, Fig. 3b).

The Earth's outer core contains around 8 wt.% light elements, such as Si, O, S, and C[41]. Si, S, and O are proposed to be the most likely major light element(s) based on recent studies[42,43]. Each weight percent of Si, S, and O light elements could reduce the thermal conductivity by 2 − 4% near CMB conditions in recent calculations and high-pressure experiments[7,21,23,44]. Thus, an additional 20 − 30% decrease in the thermal conductivity is reasonable by the light element impurities of 8 − 10 wt% for an Fe-Si-O/Fe-S-O outer core. Consequently, the thermal conductivity for liquid Fe alloy at the CMB conditions ($\kappa_{CMB}$) would be approximately 70 ± 10 W/m/K, which is 30 − 50% smaller than previous DFT computations (~100 − 140 W/m/K)[7,8]. When taking a CMB temperature of ~3,900 ± 200 K by recent studies on a Si-rich Fe alloy[45], an isentropic heat conduction down the outer core adiabat would be further constrained to be ~10 ± 1 TW (See

Methods), which contributes ~22% in the Earth's global interior heat loss (46 ± 3 TW)[13]. The determined core adiabatic heat flow of ~10 ± 1 TW near the topmost outer core in this study is comparable to the recent heat flow estimates (7 − 13 TW) from the lowermost mantle silicates across the CMB[13,46-48]. According to a recent model of heat budget in the core[20], our estimated CMB thermal heat flux of ~10 TW can be mainly contributed from a secular cooling associated with the heat capacity of the core (~4.8 TW), the latent heat associated with the freezing of the inner core (~3.3 TW), and the gravitational energy associated with the light element partitioning across the inner-core boundary (ICB) (~2.0 TW) (Fig. 4 , Methods). That is, both thermal and compositional convections play an equally important role in driving the present-day geodynamo.

Based on the recent modeling of the core thermal evolution[1,2] and the thermal heat flux of 10 TW across the CMB, the age of the inner core is constrained to be around 1.0 − 1.3 Gyr. This is significantly lower than some estimates of 3.5–4.2 Gyr[4,15], but higher than recent claims of less than ~0.7 Gyr[3]. Our study points to a 30–50% reduction in the thermal conductivity of iron alloy as compared with previous studies (~100 − 140 W/m/K) at near CMB conditions[2,3,7,8], which could translate into a difference in the estimated inner-core age as large as a factor of two. We should note that the estimation of the inner core age also depends on the thermal conductivity of the lowest mantle materials and radioactivity in the core[20]. An increase in both average field strength and variability of the Earth's palaeomagnetic field was observed to occur between 1.0 to 1.5 billion years ago in recent studies[49]. Our results would provide an explanation for the change of the observed palaeomagnetic field by the nucleation of the Earth's inner core. However, a recent magnetic evidence from samples of the ~565 million years old Sept-Îles intrusive suite shows an anomalous palaeomagnetic field during the Ediacaran period[50], indicating a young inner core of around 0.5–0.7 Gyr. The transport properties of Fe and Fe alloy in this study does not support the very young inner core nucleation. Therefore, further interrogations between mineral physics, geodynamics, paleomagnetism are needed to resolve this discrepancy.

Earth's geodynamo has maintained a magnetic field for at least 3.45 Gyr[14]. Thermal convection would play a key role to drive the early geodynamo because the inner core was smaller or even nonexistent at earlier times such that the compositional convection was much weaker. In addition, an alternative energy source for the early core from chemical differentiation such as the exsolution of magnesium (Mg) and/or oxygen (O), has been recently proposed to be a driving force for the convection for the early geodynamo[19,51,52]. In this case, Mg/O entered the core to form an Fe alloy as the core temperature was sufficiently high during the formation of the core[53]. With the core cooling, they became supersaturated and precipitated as Mg-Si-O minerals at the CMB. This process can efficiently provide a compositional convection before the inner core formation, which would drive the early dynamo, combined with the thermal convection.

**Online Content** Methods, along with any additional Extended Data display items and Source Data, are available in the online version of the paper; references unique to these sections appear only in the online paper.

## Acknowledgements

We thank Junqing Xu, Peng Zhang, P. Driscoll, T. Becker, and Jin Liu for helpful discussions. We thank N.P. Salke, J.C. Liu, Y.G. Wang, Q. Zhang, and H. Yang for their assistances with experiments. We acknowledge Y.P. Yang and Z.L. Fan for the FIB cutting. We thank X.Y. Du for the analyses of the temperature distribution. We appreciate Wenge Yang and Rossi Paul for the use of electrical resistivity measurement system. Y.Z. acknowledges supports from the National Natural Science Foundation of China (NSFC) (Grant No. 41804082) and the Fundamental Research Funds for the central universities in China. M.H. acknowledges supports from NSFC (Grant No. 41174071). J.F.L. acknowledges supports from NSF Geophysics and CSEDI Programs. We acknowledge HPsynC, Geophysical Laboratory, Carnegie Institution of Washington for the use of the Keithley ultra-low voltage (2182A model) and current source (6221 model) in the resistivity experiments. Portions of this work (*in situ* X-ray diffraction measurements and laser-heating) were performed at GeoSoilEnviroCARS (GSECARS), Advanced Photon Source (APS), Argonne National Laboratory. GSECARS is supported by the National Science Foundation - Earth Sciences (EAR-1634415) and Department of Energy-GeoSciences (DE-FG02-94ER14466). This research used resources of the Advanced Photon Source, a U.S. Department of Energy (DOE) Office of Science User Facility operated for the DOE Office of Science by Argonne National Laboratory under Contract No. DE-AC02-06CH11357. R.E.C. was supported by the European Research Council Advanced Grant ToMCaT and by the Carnegie Institution for Science. REC gratefully acknowledges the Gauss Centre for Supercomputing e.V. (www.gauss-

centre.eu) for funding this project by providing computing time on the GCS Supercomputer SuperMUC at Leibniz Supercomputing Centre (LRZ, www.lrz.de).

## Methods

**Electrical resistance measurements at high *P-T* conditions.** We used short symmetric DACs with beveled diamond anvils of 75-300 μm, 100-300 μm, and 150-300 μm culets (inner and outer culet sizes with a beveled angle of 9 degrees) to generate high pressures. Re gaskets were pre-indented to ~25 GPa and holes were drilled by laser ablation. Cubic boron nitride (cBN) was loaded into the drilled holes as a gasket insert as well as an electrical insulation, and was compressed to approximately 25 GPa. An additional hole of ~60 μm was then drilled on the compressed cBN insert and used as the sample chamber. We used polycrystalline iron sample (>99.9% purity, purchased from Alfa Aesar) as the starting sample for the electrical resistivity measurements of *hcp*-Fe at high P-T in laser-heated DACs. The sample had a thickness of ~2 μm compressed by a DAC with 600 μm culet, the same as that used in Ref. 54. We used a focused ion beam (FIB) system (FEI VERSA 3D type) to shape the sample into a uniform and suitable geometry to match the anvil culet size and the laser spot at the Center for High Pressure Science and Technology Advanced Research (HPSTAR), Shanghai, China. We used an ion beam current (Ga+) of 15 nA to cut the sample. The Fe sample had a Greek cross shape with four probes and a disc sample with a diameter of ~6 μm in the center (Fig. 1). Four Pt electrical leads were connected with each end of the Fe cross. The sample was loaded into the sample chamber and sandwiched between two dried $SiO_2$ layers with a thickness of ~5 μm each, which were used as the pressure medium and thermal insulator. The $SiO_2$ powder was fired using a furnace at ~1,400 K for ~10 hours to remove bound hydrogen before use. The use of cubic boron nitride with a high strength helped maintain the thickness of the sample chamber. Furthermore, the sample chambers were loaded with an appropriate amount of sample and pressure medium such that the sample was evenly compressed at high pressure and there's minimal shear stress induced deformation in the chamber. Extended Data Figure 9 shows that the sample was evenly compressed and maintained its shape up to ~140 GPa. The Fe sample and $SiO_2$ were loaded in a clean operating lab within a relative humidity of 30-40% to keep the sample chamber dry to minimize adsorbed water during the experimental preparation. The chemical composition of iron wire was analyzed before and after the experiments using energy dispersive spectroscopy and X-ray diffraction, which did not show chemical contaminations by Pt leads or $SiO_2$ thermal insulator.

Five sets of experiments at 82(2) GPa, 105(2) GPa, 133(2) GPa, 142(3) GPa, and 165(3) GPa (pressures measured using X-ray diffraction of the sample at ambient temperature) were conducted. The total pressures given in Fig. 2 and Extended Data Figs. 3 and 6 are those at room temperature. The pressure was calculated from the measured lattice parameter of *hcp*-Fe by *in site* XRD and their thermal equation of state[32] (Extended Data Fig. 4). The samples were continuously heated from both sides using a double-sided laser heating system with a focused laser beam spot size of ~10 μm at the 13ID-D station at the GeoSoilEnviroConsortium for Advanced Radiation Sources (GSECARS) of the Advanced Photon Source (APS), Argonne National Laboratory (ANL). We used two flat-top laser beams to homogeneously heat the sample from both sides, which helped to minimize the temperature gradient in the heated area[31]. The sample temperature was measured by fitting measured thermal radiation spectra to a Plank function. The temperature variation in the flat-top laser heating spot within ~8 μm is approximately ≤5% (approximately 50 − 150 K) from ~1,300 K to ~3,000 K (Supplementary Information and Extended data Fig. 5)[31]. In our experiments, we ensured that the Fe sample formed a circular disk of approximately 6 μm in the central area of the Greek cross shape so the focused laser spot was larger than our sample size (~6 μm), which guaranteed our sample in a homogeneous heating status. Optical observation of the sample sandwiched between silica layers was used to ensure that the sample shape was maintained to permit homogeneous laser heating at high *P-T* and to permit reliable four-probe electrical conductivity measurements, instead of pseudo-four probe measurements as in some previous studies (ref. 3).

A constant direct current of 5 mA was applied onto the Fe sample using a Multimeter source (Keithley 6221 model). The voltage of the Fe sample at high *P-T* was measured using an ultra-low voltmeter (Keithley 2182A model). The electrical resistances were then obtained by Ohm's law. When the direct current passed through the sample from leads A to B, the voltage ($V_1$) was measured between probes C and D as shown in Fig. 1. Meanwhile, when the current went from B to C, the voltage ($V_2$) was also measured between A and D. We used the average value of measured voltages (($V_1 + V_2$)/2) to obtain the resistance. *In situ* synchrotron XRD spectra were also collected by a Pilatus 1M CdTe detector before, during, and after the laser heating, respectively, using an incident X-ray beam of ~3 μm in diameter (FWHM) and 0.3344 Å in wavelength (Extended Data Fig. 1).

**Electrical resistivity and Bloch-Grüneisen formula at high *P-T*.** The electrical resistivity ($\rho$) at high temperature is derived from the measured resistance ($R$) and sample volume ($V$), as well as the resistance ($R_0$) and sample volume ($V_0$) at room temperature. The sample was a disc and was visible optically during our experiments, so we could assume the sample geometry changes isotropically at high *P-T*, and the resistivity can be obtained by[2]:

$$\frac{\rho}{\rho_0} = \frac{R}{R_0}\frac{l}{l_0} = \frac{R}{R_0}\left(\frac{V}{V_0}\right)^{1/3} \quad (1)$$

where $l_0$ and $l$ are the thickness of the sample at ambient conditions and at high P-T, respectively; $l/l_0$ is equal to $(V/V_0)^{1/3}$ if the geometry change of the iron sample is isotropic; $\rho_0$ is the measured resistivity of *hcp*-Fe at high pressure and room temperature (Extended Data Figure 2). Two independent resistivities $\rho_1$ and $\rho_2$ were obtained in relation to the measured resistances $R_1$ and $R_2$ with increasing temperature at ~142 GPa when the current went through AB and CD paths, respectively (Extended Data Fig. 10). The averaged resistivity in these measurements had an uncertainty of about 4–5 %. It should be noted that the uncertainties of our reported resistances are propagated from the measured pairs only. Uncertainties from the cases of anisotropic deformation had also been considered, including the geometry change along the compression axis or in the radial direction[2]. If the sample geometry changes along the compression axis only, $l/l_0$ is equal to $V/V_0$, which will give a lower bound for the measured resistivity. On the other hand, if the sample geometry changes along the radial direction only, then $l$ is equal to $l_0$, which will give an upper bound for the measured resistivity. For example, $V/V_0$ increases from 1.000 to ~1.005 with the temperature increasing from ambient temperature (15.833(59) Å$^3$) to 2376(120) K (15.908(25) Å$^3$) at ~142 GPa in Extended Data Fig. 1b. Therefore, the two extreme anisotropic deformations will give an uncertainty of less than 1% in the thickness of *hcp*-Fe sample at high P-T. The total uncertainty in the measured resistivity has been estimated through the standard propagation of uncertainty in Eq. (1):

$$\Delta\rho = \left[\left(\frac{\partial\rho}{\partial\rho_0}\Delta\rho_0\right)^2 + \left(\frac{\partial\rho}{\partial R_0}\Delta R_0\right)^2 + \left(\frac{\partial\rho}{\partial R}\Delta R\right)^2 + \left(\frac{\partial\rho}{\partial l_0}\Delta l_0\right)^2 + \left(\frac{\partial\rho}{\partial l}\Delta l\right)^2\right]^{1/2} \quad (2)$$

The measured resistivity at high pressure and room temperature ($\rho_0$) has an uncertainty of around 5%. After standard error propagations, the estimated one sigma error ($\pm 1\sigma$) is approximately 8% for the resistivity of *hcp*-Fe at relevant P-T conditions of the core. These uncertainties of the measured resistivity are plotted in Fig. 2 and Extended Data Fig. 6.

The measured resistivity increased quasi-linearly with increasing temperature up to ~3,000 K at high pressures, which can be described using the Bloch-Grüneisen formula:

$$\rho_{BG}(V,T) = D(V)\left(\frac{T}{\theta_D(V)}\right)^n \int_0^{\theta_D(V)/T} \left[\frac{z^n}{(e^z-1)(1-e^{-z})}\right] dz \qquad (3)$$

where the $V$ can be obtained from *in site* XRD, Debye temperature $\theta_D(V)$ is from the previous study[55], constant $n$ and $D(V)$ could be yielded through fitting the measured resistivity and temperature at each pressure point.

**Isentropic (adiabatic) heat flow across the core-mantle boundary.** Isentropic heat flow along temperature gradient was given by[56]:

$$Q_s(r) = -\oint q \cdot dS = -4\pi r^2 \kappa \left(\frac{\partial T}{\partial r}\right)_S \qquad (4)$$

where $\kappa$ is thermal conductivity, $r$ is the radius, $S$ is the surface area, $q$ is heat flux across the temperature gradient $\frac{\partial T}{\partial r}$. For an isentropic temperature gradient across the CMB, the heat flow is:

$$Q_{CMB} = 4\pi r^2 \kappa_{CMB} \left(\frac{g\gamma T}{\Phi}\right)_{CMB} \qquad (5)$$

where $\kappa_{CMB}$ is the thermal conductivity of the core materials at CMB conditions, $g$ is gravity acceleration (10.6823 m/s$^2$)[57], $\Phi$ seismic parameter (65.05 km$^2$/s$^2$ from the PREM[57]), $\gamma$ is the Grüneisen parameter (~1.5)[42] at the CMB, $T_{CMB}$ is ~3,900 K taken from the determined melting temperature of Fe-Si-O alloy at near CMB conditions[45].

**The energy balance of the present core.** The energy budget in the Earth's core may be written as[20]:

$$Q_{CMB} = Q_S + Q_L + Q_G + Q_{rad} \qquad (6)$$

where $Q_S$ is core secular cooling of the core, $Q_L$ is latent heat delivered as the inner core freezing, $Q_G$ is gravitational energy associated with the release of the light element as the inner core solidification, and $Q_{rad}$ is heat from the decay of radioactive elements within the core. The core secular cooling is

$$Q_S = -M_C C_C \dot{T}_C \qquad (7)$$

where $M_C$ is the core mass, $C_C$ is core specific heat capacity, and $\dot{T}_C$ is the present-day change rate of the core temperature. The core makes up 32.5% of the planet's total mass ($M_\oplus$). We took the estimated specific heat capacity ($C_C$) of ~800 J K$^{-1}$ kg$^{-1}$ in the core[8,20]. $\dot{T}_C$ can be estimated from the CMB temperature change from the ancient to present-day core[20]. The $Q_L$ and $Q_G$ is related to the growth rate of the inner core:

$$Q_L = \dot{M}_{IC} L \tag{8}$$

$$Q_G = \dot{M}_{IC} E_G \tag{9}$$

where $\dot{M}_{IC}$ is the mass change of the inner core $M_{IC}$, $L$ and $E_G$ are the latent and gravitational energy per unit mass. The $L$ is around ~500–600 kJ/kg[58], the density at top of the inner core[57] is $12.8 \times 10^3$ kg/m³, and $E_G$ is ~$3 \times 10^5$ J/kg[20] at the ICB conditions. The core growth rate is estimated to be ~0.9 mm/year from the determined inner core age of 1.2 − 1.3 Gyr in this study. The details for the used parameters are listed in Extended Data Table 2. Finally, we can obtain the $Q_S$ = ~4.8 TW, $Q_L$ = ~3.3 TW, and $Q_G$ = ~2.0 TW, respectively. The radiogenic heat by the decay of $^{40}$K is estimated range from 0.2 to 1.9 TW[59] due to the not well determined $^{40}$K concentration in the present core.

**Figures**

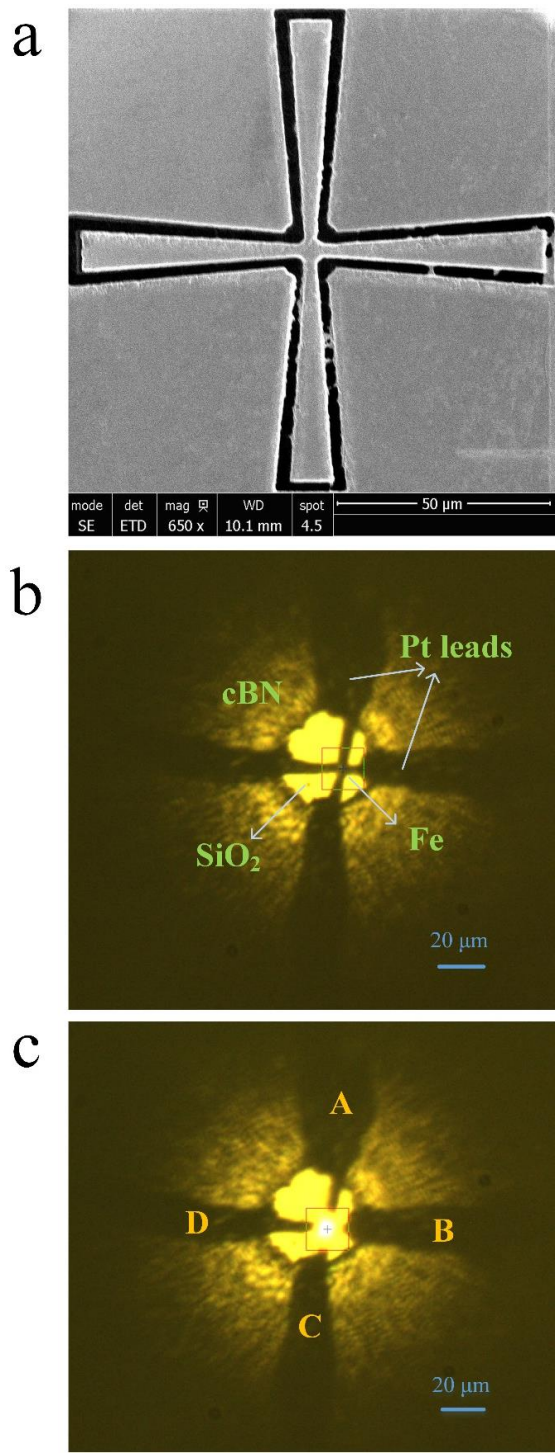

**Figure 1 | Photograph of a shaped iron foil and the sample loaded in a laser-heated diamond anvil cell (DAC).** **(a)** A Greek cross sheet of iron shaped by FIB; and **(b)** a loaded sample in a DAC with a culet of 300 μm beveled to 75 μm at ~142 GPa and 300 K and then laser heated

to 2,400 K (**c**). Pt leads had a width of ~20 μm and a thickness of ~4 μm. The laser was focused into a flat-top shape on the circular section of the Fe sample with approximately 6 μm in diameter.

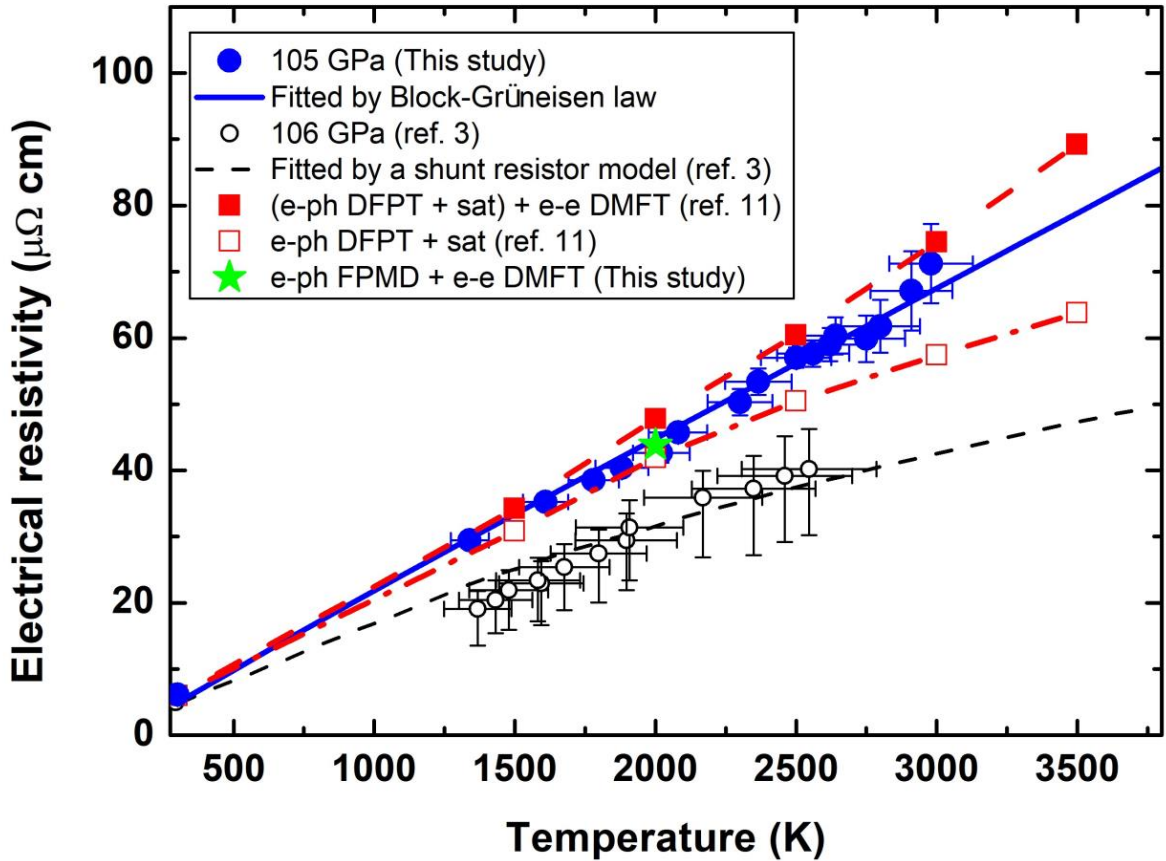

**Figure 2 | Measured and calculated resistivity of *hcp*-Fe at ~105 GPa with increasing temperature up to ~2,980 K in a laser-heated DAC.** The pressure of 105 GPa was determined at ambient temperature by X-ray diffraction. At ~2,980 K, the pressure is about 117 GPa according to the thermal equation of state of *hcp*-Fe[32]. Our measurements are compared with the current calculations (solid squares) that considered both "*e-ph*" and "*e-e*" contributions by "DFPT + DMFT" (ref. 11) and new calculations (solid green star) by "FPMD + DMFT" (this study), and also compared with previously measured resistivity[3] (open circles) and calculations considered only "*e-ph*" contribution (ref. 11, open red squares). "*e-ph*" is the electron-phonon contribution of resistivity calculated by DFPT with inelastic Boltzmann theory, "*e-e*" is the electron-electron contribution of resistivity (open squares), and "sat" is resistivity saturation effects for the "*e-ph*" scattering. The red dashed line through the calculated points in the resistivity with increasing temperatures is guides to the eye. The solid blue curve is fitted using the Bloch-Grüneisen formula for the measured resistivity. The black short dashed line represents the saturated resistivity fitted by a shunt resistor model in Ref. 3.

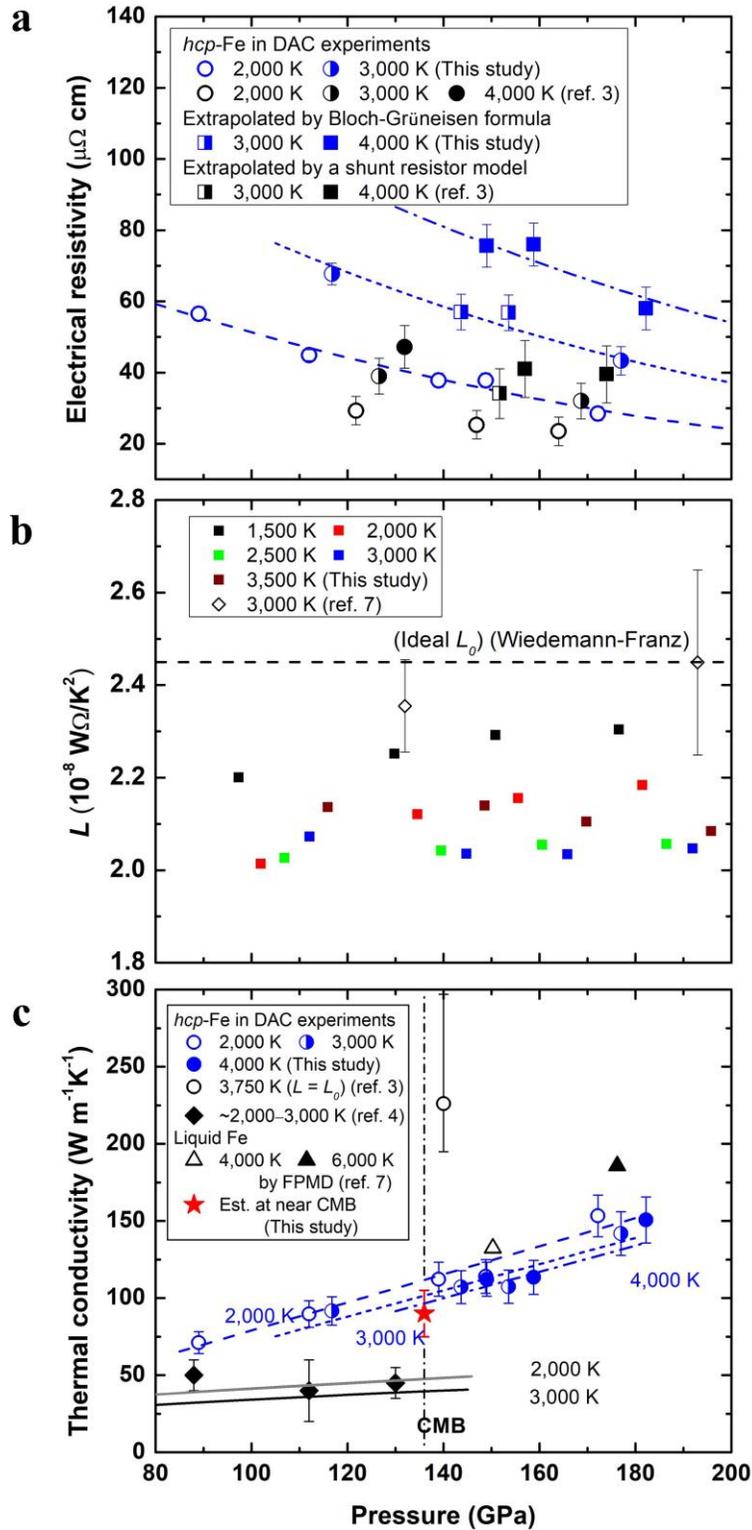

**Figure 3 | Electrical resistivity and thermal conductivity of Fe at the relevant *P-T* conditions of Earth's core.** (**a**) Our experimental results of *hcp*-Fe at 2,000 K (open blue circles) and 3,000 K (semi-open blue circles), respectively, and the extrapolated data by Bloch-Grüneisen

formula at 3,000 K (semi-open blue squares) and 4,000 K (solid blue squares) at high pressures, respectively. Our measured resistivities are compared with previous results (ref. 3) at 2,000 K (open black circles), 3,000 K (semi-open black circles), and 4,000 K (solid black circles), respectively, and their extrapolated data by resistivity saturation model at 3,000 K (semi-open black squares) and 4,000 K (solid black squares), respectively. (**b**) Calculated Lorenz number (*L*) as functions of pressure and temperature. Lorenz number via the Wiedemann-Franz (W-F) relation that is derived from the ref. 3, and the ideal Lorenz number $L_0 = 2.44 \times 10^{-8}$ W$\Omega$/k$^2$. Errors in our calculated Lorenz numbers (~2%) are small than the symbol size. The calculated Lorenz number in liquid Fe (ref. 7) is also plotted as comparisons (open diamonds). (**c**) The thermal conductivity of *hcp*-Fe and liquid Fe derived using the Wiedemann-Franz relation at the relevant conditions of the outer core. The thermal conductivities are compared with previous results derived from resistivity experiments (open black circle, ref. 3) and from monitoring a heat pulse propagation (solid black diamonds, ref. 4), respectively. The star represents the derived thermal conductivity of liquid Fe in this work at near CMB conditions (~136 GPa and ~4,000 K), which is compared with the calculated results in liquid Fe[7] at 4,000 K (open black triangle) and 6,000 K (solid black triangle), respectively. The lines through the experimental points are guides to the eye, where dash, short-dash, and dash-dot lines represent the electrical resistivity and thermal conductivity as a function of pressure at 2,000 K, 3,000 K, and 4,000 K, respectively. The vertical dash-dot line represents the core-mantle boundary.

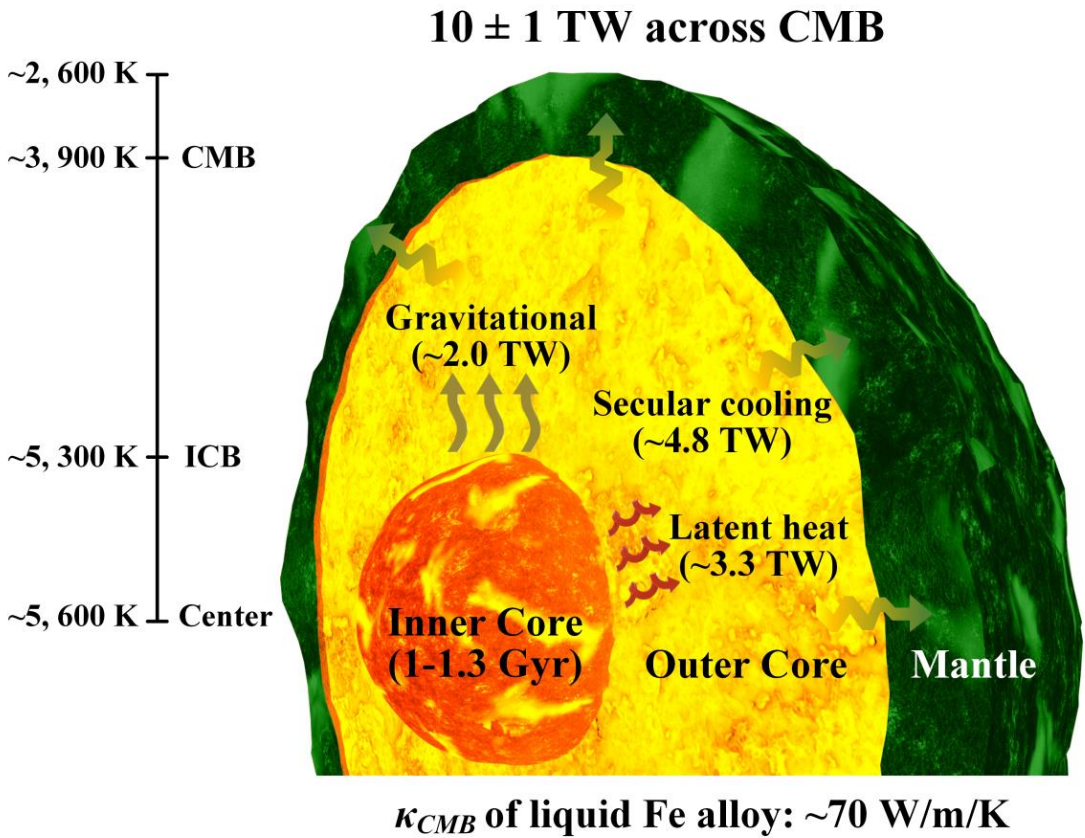

**Figure 4 | Isentropic heat conduction down the outer core adiabat and the energy balance of the present-day core.** The present-day geodynamo is driven by both thermal and compositional convections with an isentropic heat flow of ~10 TW across the CMB, including core secular cooling of ~4.8 TW, as well as latent heat of ~3.3 TW and gravitational energy of ~2.0 TW at ICB. $\kappa_{CMB}$ is the thermal conductivity of liquid Fe alloy at the relevant conditions of the CMB. The current inner core occupies ~4.3 volume % of the core and is estimated to be 1.0–1.3 Gyr old. As it grows larger in the future, the compositional convection is expected to contribute more significantly to power the geodynamo.

# Supplementary Information

## SI Text:

**Effect of temperature gradient on the electrical resistivity measurements.** A test run at ~74 GPa was conducted to examine the temperature gradient effect on the electrical resistivity measurements of *hcp*-Fe. A Greek cross Fe sample was prepared by FIB. It had a circular central area with ~15 μm in diameter as the sample, which is purposely prepared to be bigger than the laser spot (~10 μm). As shown in Extended Data Fig. 3a (red open circle), we intentionally heated the sample on one corner of the circular area (such as the corner between leads A and B), which could cause a temperature gradient from the corner AB to CD. A schematic diagram at the top-right Extended Data Fig. 3a shows the geometry of four electrodes viewed along the compression axis of the sample chamber and the temperature distribution across the cross-heart area from analysis of the thermal radiation spectra. The reported temperature was the maximum temperature ($T_{max}$) from the spectral radiation measurements. The area close to leads A and B was heated to the $T_{max}$, which was higher than the area close to the corner between leads C and D. Then we measured electrical resistivity when a direct current went through leads from A to B and from C to D, respectively. As a result, two sets of data were collected up to ~2,700 K at ~74 GPa. The resistivity when the current went through the AB path ($\rho_{AB}$) was much higher than that the one through the CD path ($\rho_{CD}$) because of the temperature gradient. The $\rho_{AB}$ increases generally quasi-linearly with increasing temperature, which could be fitted by the Bloch-Grüneisen formula (solid blue curve, Extended Data Fig. 3b). The $\rho_{AB}$ is overall consistent with the previous results fitted from the data at 75 GPa and temperature from 300 K to 450 K in a muffle furnace (solid black curve[3], Extended Data Fig. 3b). The $\rho_{CD}$ is close to the data at ~80 GPa by Ohta et al. (2016), which looks depressed by "high temperature" and could be fitted by a shunt resistor model with a saturation resistivity of ~82 μΩ cm. In the latter case, the current went through colder regions but did not go through the real high-temperature region[5], so its resistivity was depressed by a false "high temperature". Our results even show that the $\rho_{AB}$ could be around two times higher than the $\rho_{CD}$ up to ~3,000 K at 75 GPa in Extended Data Fig. 3b. We, therefore, conclude that the temperature-induced strong resistivity saturation observed previously in *hcp*-Fe is most likely caused by an artificial temperature gradient as a result of the sample geometry and inhomogeneous heating.

**Possible texture effects on the electrical resistivity of *hcp*-Fe.** Texturing effects on the electrical conductivity anisotropy could also affect the interpretation of our results. Here we consider the following three factors:

**(1). Deviatoric stress and shear stress induced lattice-preferred orientation:** Under non-hydrostatic compression in DAC, polycrystalline *hcp*-Fe can develop deviatoric stress-induced lattice preferred orientations (textures) with the *c* axes parallel to the compression axis of the DAC[60-62]. Such textures can also be developed under shear stress deformation when *hcp*-Fe crystals were deformed unevenly (shear induced sample flow). Ohta et al.'s samples in 2016 and 2018 exhibited strong textures likely due to both effects. In this study, our experiments were designed to reduce these effects on texture developments. We used $SiO_2$ as a pressure transmitting medium and annealed Fe samples at ~1,500 K for a few minutes at increasing pressures to reduce the deviatoric stress in the sample chamber. Cubic boron nitride was used as the gasket insert to increase the thickness and strength of the sample chamber. Furthermore, the sample chambers were loaded with an appropriate amount of sample and pressure medium such that the sample was evenly compressed at high pressure and there's minimal shear stress induced deformation in the chamber. We added Extended Data Figure 9 to show that the sample was evenly compressed and maintained its shape up to ~140 GPa. Analysis of the X-ray diffraction patterns of our *hcp*-Fe samples taken along the compression axis of the DAC at high pressures shows an intensity ratio between (100) and (002) peak of ~2:1 (Extended Data Figs. 1a and 1b). If strong textures were to be developed, one would expect basal planes to be perpendicular to the compression axis and thus very minimal intensity of the (002) peak (see Wenk et al. (2000) for details). Therefore, the degree of texturing in our *hcp*-Fe samples is likely in between non-hydrostatic compression in Wenk et al.'s study and hydrostatic compression in typical equation of state experiments.

**(2). High temperature-induced recrystallization:** The grain size and orientation of polycrystalline *hcp*-Fe can change during laser heating especially when the temperature is close to the melting point. This has been observed in previous studies and is called fast recrystallization[63]. In our experiment, we only heated iron samples up to ~3,000 K, which were below the fast recrystallization threshold (Extended Data Fig. 4). Analysis of X-ray diffraction patterns at high P-T shows no significant spotty patterns from crystal coarsening (Extended Data Fig. 1a), and the measured electrical resistivity increases linearly with temperature.

**(3). Theoretical calculations:** First-principles results also show an electrical resistivity anisotropy of $\rho_a/\rho_c = 1.3$ in *hcp*-Fe at Earth's core conditions[11]. Ohta et al. (2018) reported approximately 30% anisotropy in modeled thermal conductivity in various Fe samples (powder, foil, and wire iron with different degrees of textures) at 20–45 GPa and 300 K using laser pump-and-probe measurements[36]. We should note that the experimental methodology in Ohta et al.'s study typically has an uncertainty in the order of tens of percent. Most importantly, Ohta et al. claimed a very large anisotropy of $k_c/k_a = 3$–4 through extrapolation of limited data (at 20–45 GPa and 300 K) to Earth's core P-T conditions[36]. The uncertainties in their extended extrapolations would be too large to make the claim credible. Their measured anisotropy of a maximum 30% is consistent with the first-principles estimates.

Considering the aforementioned results with anisotropy of 30% and texture between non-hydrostatic and hydrostatic conditions, the thermal conductivity anisotropy of *hcp*-Fe is likely in the order of ten percent maximum. This is comparable to the range of our reported uncertainties for the electrical measurements so it does not affect our conclusions. It also could not account for the discrepancy between Ohta et al. (2016), Konopkova et al. (2016), and this study.

**First-principles molecular dynamics (FPMD) computations of transport in *hcp*-Fe.** We performed classical molecular dynamics, integrating F = m·a, with ab initio density functional forces computed using Quantum Espresso for a density of 11.00 g/cm$^3$ at 2000 K[64]. We used the GBRV ultrasoft pseudopotential with the standard plane-wave cut-off energy of 40 Ry[65]. The time step is set to 1 fs. We run the simulations for 8 ps or longer and extract several configures separated by 1 ps. Simulations were performed in the NVT ensemble for longer than 10 ps at atomic volume 56.5 bohr[64] and temperature of 4,000 K. We apply DFT-DMFT and Kubo formula to calculate transport properties of selected snapshots and average them to obtain the final results. For transport calculations, k-point grids 4×4×2 are used. The number of snapshots is enough to converge the transport properties to better than 2%. We studied different sized cells and k-point sets to estimate finite size errors, and found that with Γ point sampling for at least 288 atom supercells are required (6×6×4 *hcp* cells). The result is shown in Extended Data Fig. 7.

**Uncertainty in the first-principles computations.** Internally we checked the convergence of our results with respect to all controllable parameters including system size, k-points, run time, q-points, etc., as appropriate for each method. Beyond this, the best way to assess the accuracy of transport computations is by comparing with experiment, as we did in this paper. The estimates of the saturation effects and the neglect of the anharmonicity may introduce errors of up to 10 − 15% for the *e-ph* contribution. Uncertainty of the Hubbard U parameter could lead to some uncertainties of the *e-e* contribution. In total, the uncertainties of total resistivity and thermal conductivity calculated using linear response are about 15% − 20% at most. For the method employing first-principles molecular dynamics to estimate melting effects, uncertainties may be about 10 − 15% due to the finite size issues, and the chosen of the Hubbard U parameter. The total uncertainties are probably as large as compositional effects in the outer core, which are presently unknown.

**Electrical resistivities of *hcp*-Fe at high *P-T* between experiments and first-principles computations.** We compare more measured resistivities with the current calculations (ref. 11) at high P-T. At ~142 GPa (Extended Data Fig. 6c), the experimental and theoretical resistivity also agrees. At ~165 GPa (Extended Data Fig. 6d), the calculated resistivities are systematically higher than experimental results, which is even ~35% higher than the extrapolated value from the Bloch-Grüneisen formula at ~3,500 K. Saturation is actually not a well-understood phenomenon, especially for electron-electron scattering. There is no concept of scattering being limited by the lattice constant, and even for electron-phonon scattering the parallel resistor formula has only been proved for certain simple models[66-68] and has been a problem in condensed matter physics for many decades. In fact, using the term "saturation" is even a misnomer, because the effects of short-range scattering can even lead effects of either sign, to increases in resistivity[69,70], or far beyond the Ioffe Regel limit[71,72].

**The estimates of the Lorenz number of *hcp*-Fe at high *P-T*.** The Lorenz number that relates electrical resistivity and thermal conductivity, is fairly constant for most of metals and alloys at ambient pressures, although strong violations are known[73,74]. The measured and calculated electrical resistivity of *hcp*-Fe at high temperatures as a function of pressure from ~80 to ~200 GPa diminishes gradually with increasing pressures, but the slope becomes shallower up to ~200 GPa (Extended Data Fig. 8a). The calculated thermal conductivity of *hcp*-Fe, which considered both *e-ph* and *e-e* contributions (ref. 11), overall increases with increasing pressure but decreases with increasing temperature from ~1,500 K to 3,500 K (Extended Data Fig. 8b). We thus derive the effective Lorenz number from the computed resistivity and thermal conductivity (ref. 11), and find a 10 to 20 % lower value than $L_0$ (2.44× $10^{-8}$ WΩK$^{-2}$) at high P-T (Extended Data Fig. 8c), consistent with previous calculations in liquid Fe[7]. It generally decreases with increasing temperatures, and is ~(2.0 − 2.1) × $10^{-8}$ WΩK$^{-2}$ at the temperature up to 2,500 K and above. The computed Lorenz number in *hcp*-Fe seems not strongly temperature-dependent, especially above ~2,500 K at between 80 and 200 GPa. While a previous study shows that the Lorenz number may be significantly reduced by increasing both pressure and temperature at relatively low P-T (< 6 GPa and 2,100 K) in liquid Fe[75]. Further studies are needed to ascertain the Lorenz number of liquid Fe at higher P-T conditions.

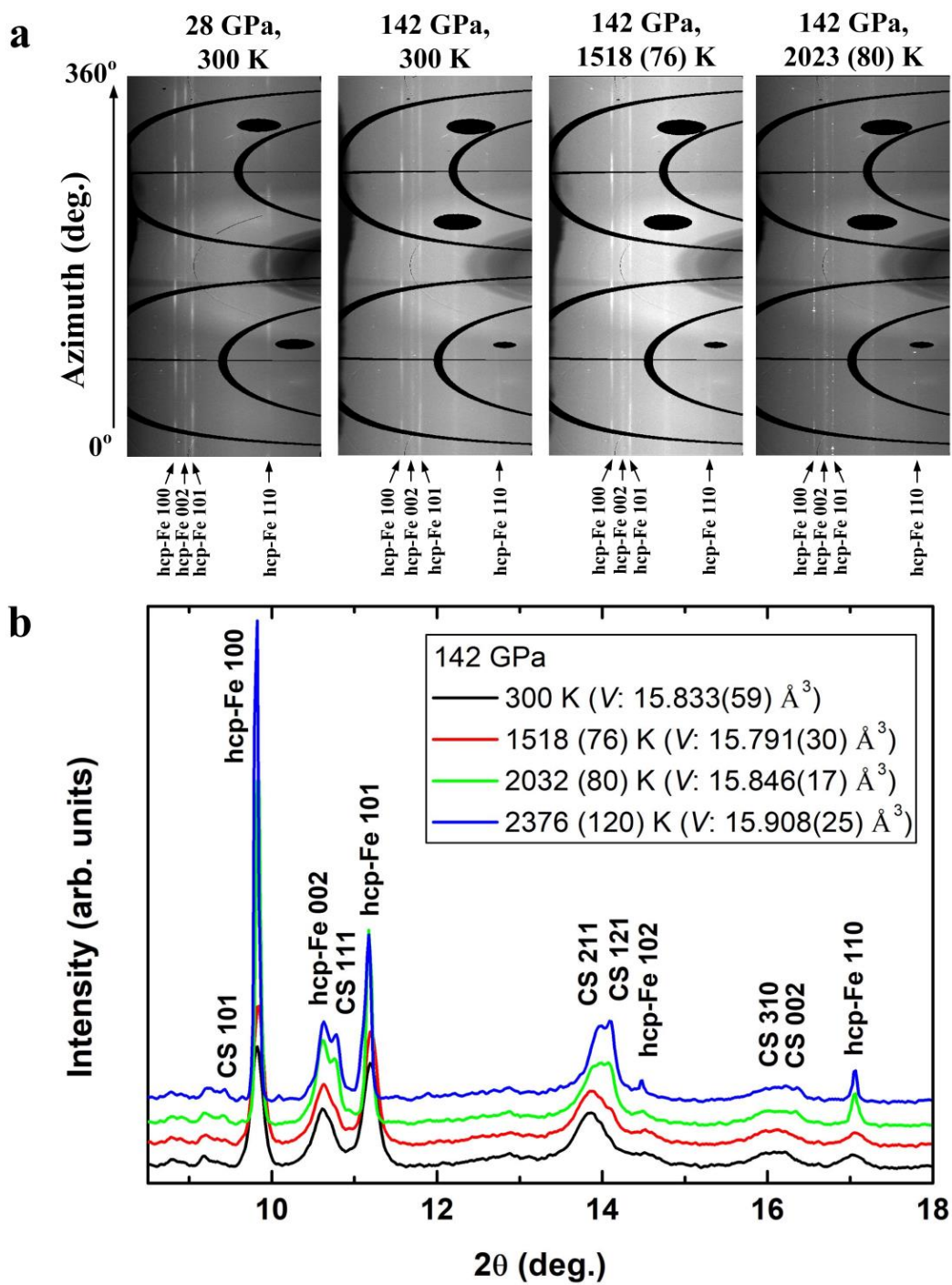

**Extended Data Figure 1 | *In situ* XRD patterns of an *hcp*-Fe sample at high pressure-temperature. (a)** Representative unrolled XRD images of *hcp*-Fe sample at ~28 GPa (300 K) and ~142 GPa (300 K, 1518 K, and 2023 K). **(b)** XRD with Miller indices of *hcp*-Fe and CaCl₂ type SiO₂ at high pressure-temperature. CS means SiO₂ component with the CaCl₂ type phase at high pressure-temperature. Data were collected at 142 GPa at ambient temperature. At high temperatures of 2032

K and 2376 K during laser heating, the pressures were about 149 GPa and 151 GPa, respectively, by the thermal equation of state of *hcp*-Fe[32]. *V* represents the measured volume of *hcp*-Fe at high P-T by *in-situ* XRD, where with increasing temperature from room temperature (300 K) to 2376(120) K the volume only increased ~0.5%.

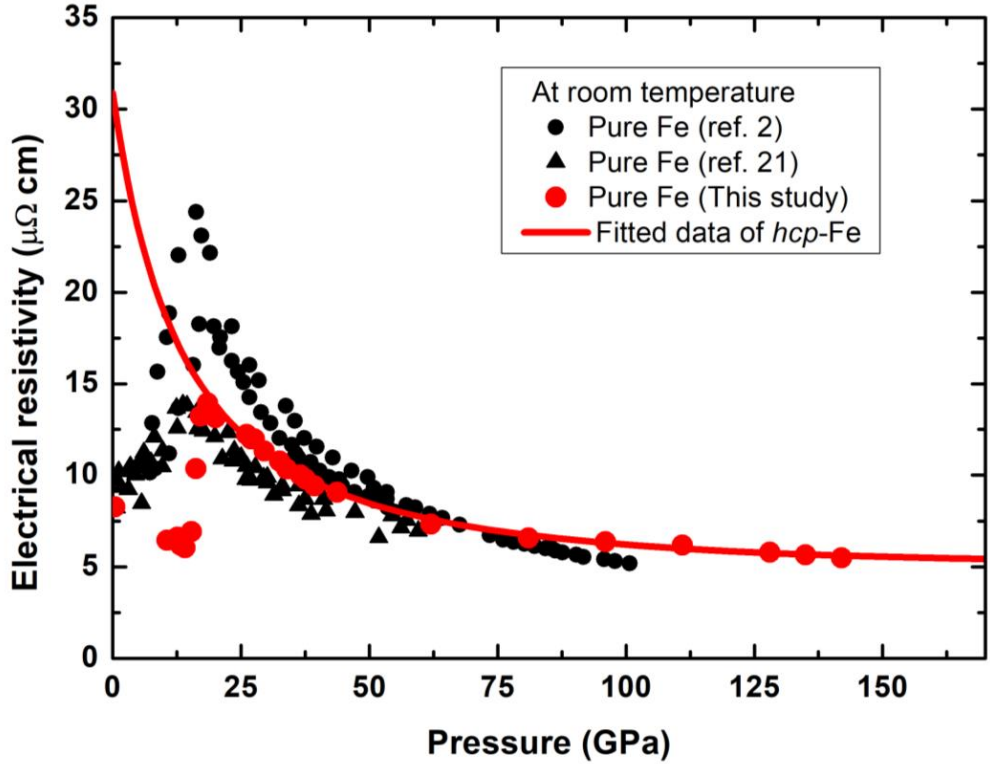

**Extended Data Figure 2 | Measured electrical resistivity of solid Fe at high pressure and room temperature.** Black circle and triangle represent the high-pressure experiments by Gomi et al. (2013)[2] and Seagle et al. (2013)[21], respectively. The red circle is this measurement together with the results of Zhang et al. (2018)[54] used the same sample. The red curve is the resistivity of *hcp*-Fe at high pressures fitted from our measurement data.

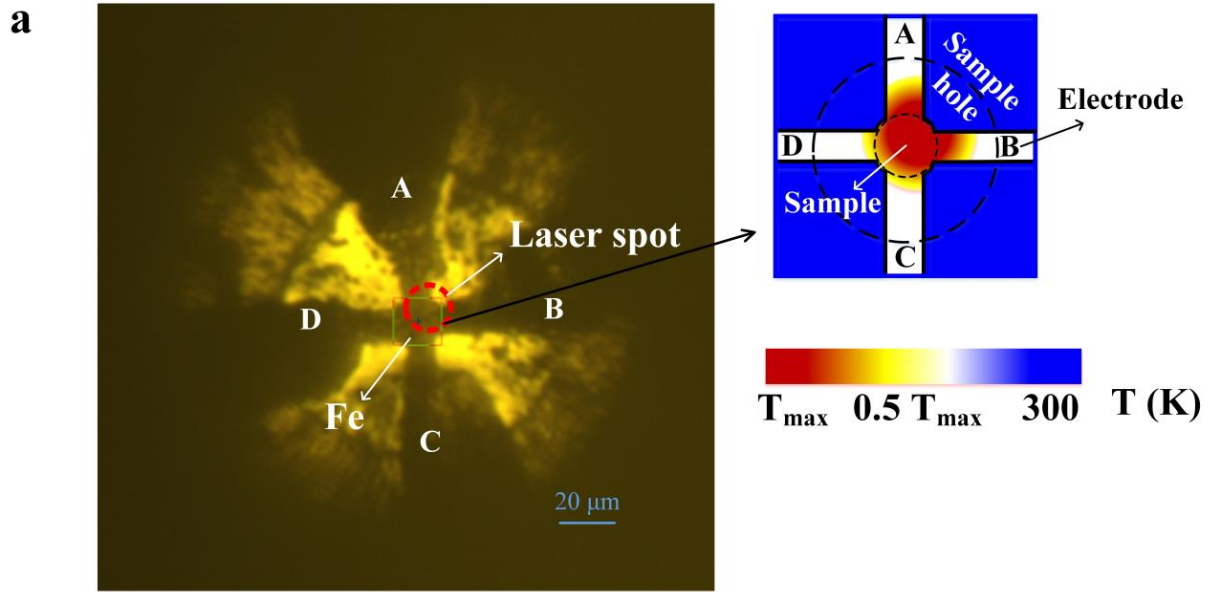

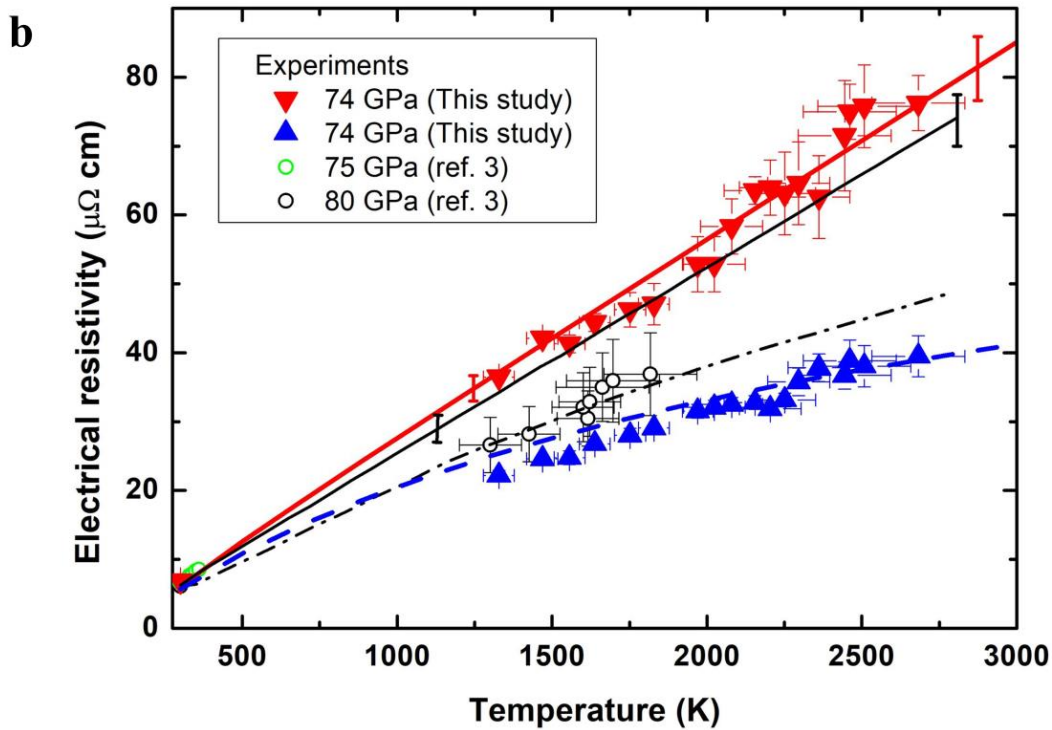

**Extended Data Figure 3 | A test run on temperature effect at 74 GPa in a laser-heated DAC.** (**a**) Microphotograph of the sample loaded in a diamond anvil cell with a culet of 200 μm. The heart of the sample with a Greek cross shape is around 15 μm in diameter, which is slightly larger than the laser spot (~10 μm as shown in the red open circle). The schematic diagram at the top right depicts the geometry of four electrodes viewed along the compression axis of the sample

chamber and the temperature distribution across the cross-heart area from analysis of the thermal radiation spectra. (**b**) The measured resistivity versus temperatures for a test run at ~74 GPa. One corner of Fe sample between electrodes A and B was heated such a temperature gradient was made artificially, where the region close to the corner AB had a higher temperature than that close to the corner CD. When the current went through from lead A to B, the measured resistivity ($\rho_{AB}$, red inverted triangles) is ~2 times higher than the one $\rho_{CD}$ (blue triangles) that went from C to D. The red and black solid curves represent the temperature-resistivity slope fitted by the Block-Grüneisen formula in this study and previous data from 300 K to 450 K in muffle furnace (ref. 3), respectively, which agree with each other overall. The blue dashed line and black dash-dotted line represent the temperature-resistivity relation fitted by a shunt resistor model in this study and the previous data (ref. 3) at ~80 GPa in a laser-heated DAC, respectively.

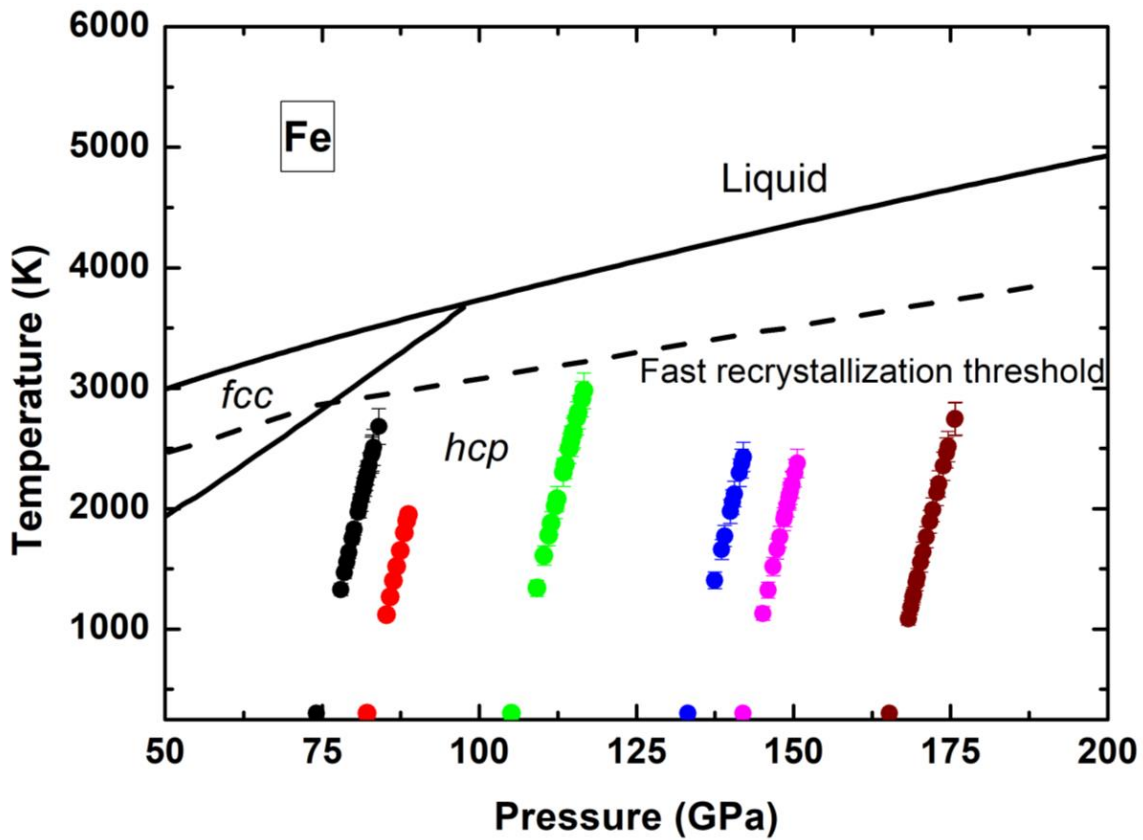

**Extended Data Figure 4 | Pressure and temperature conditions in our electrical resistivity measurements of *hcp*-Fe in this study.** Phase boundary and fast recrystallization threshold are from the literature[63]. Colourful circles represent the conditions of laser-heated DAC experiments.

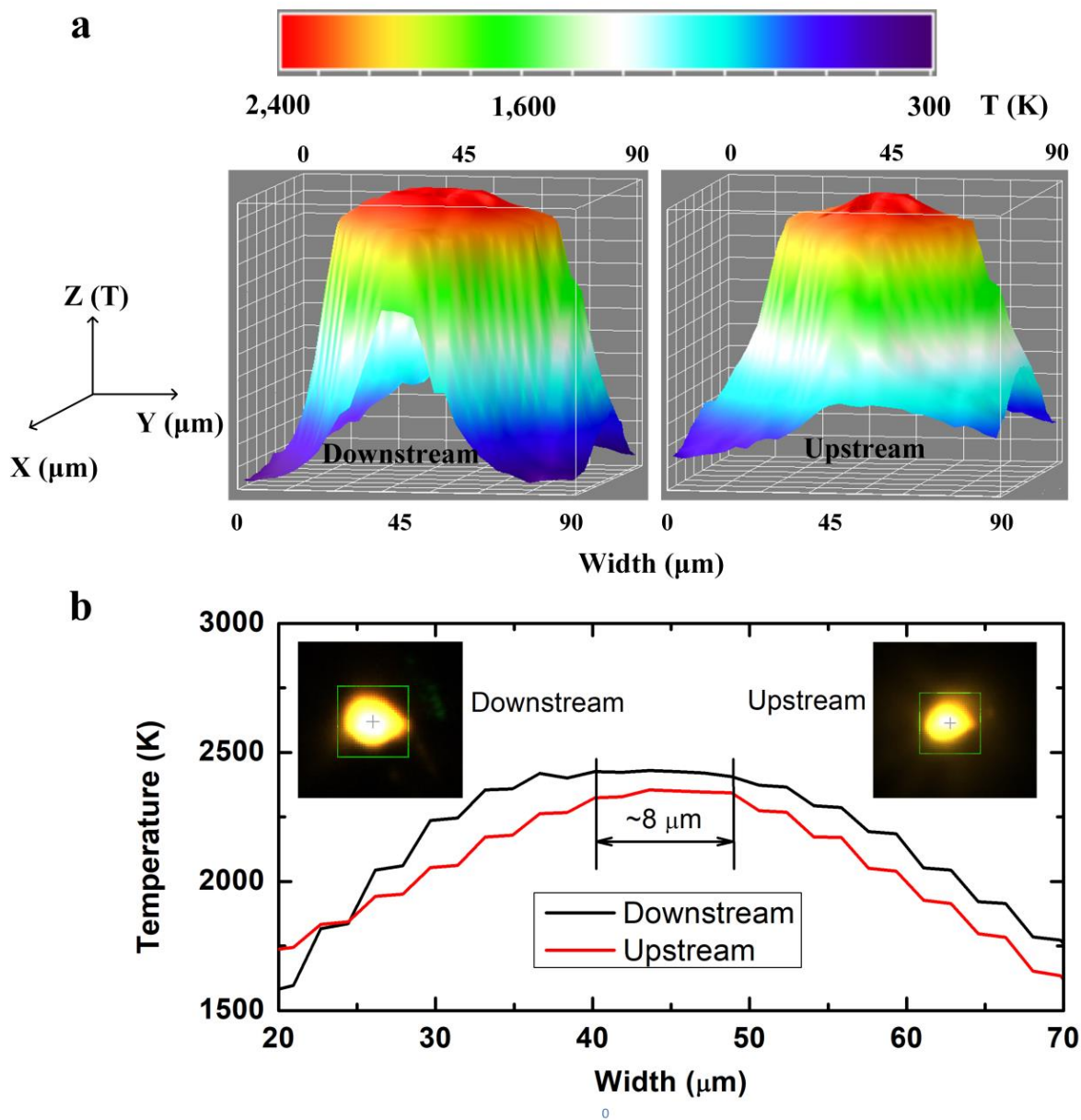

**Extended Data Figure 5 | Temperature mapping of laser-heated Fe sample at ~142 GPa and ~2,380(120) K.** (**a**) 3-D temperature distributions of heated Fe sample in downstream and upstream sides, respectively; (**b**) A temperature flat area with ~8 μm in diameter. The inserts in (b) are *in situ* microphotographs of downstream and upstream sides of the sample on laser heating. The temperature variation of Fe sample within ~8 μm for resistance measurements is ~100 K between the downstream and upstream sides as shown in (b).

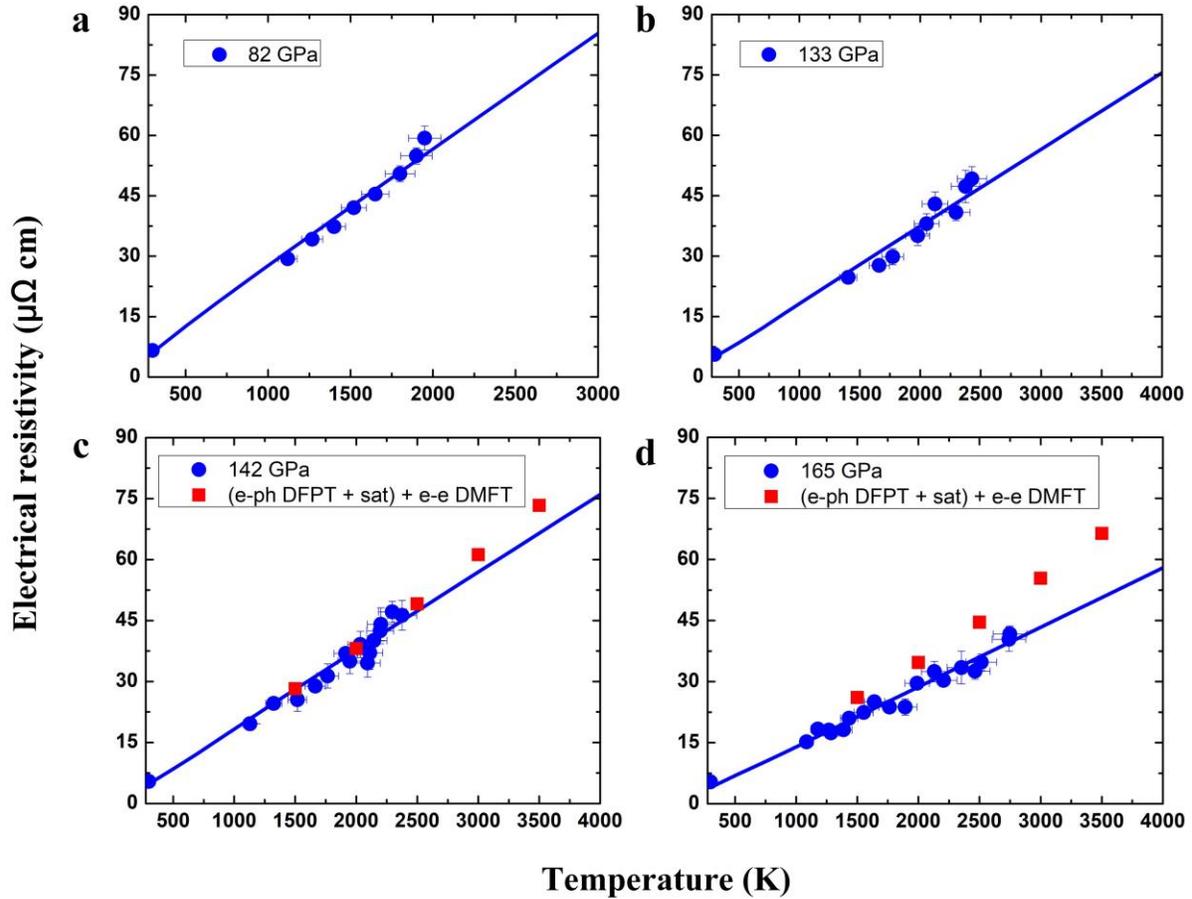

**Extended Data Figure 6 | Measured and calculated electrical resistivity of *hcp*-Fe with increasing temperature at high pressures.** Electrical resistivity measured up to ~1,950 K at ~82 GPa (**a**), up to ~ 2,430 K at ~133 GPa (**b**), up to ~2,380 K at 142 GPa (**c**), up to ~2,750 K at ~165 GPa (**d**). The pressures of 82, 133, 142, and 165 were measured at ambient temperature. The measured resistivities are compared with the recent first-principles calculations at ~142 and ~165 GPa (ref. 11), respectively. The blue curves represent the resistivity responses with temperature in *hcp*-Fe fitted by the Block-Grüneisen formula. The pressures were determined by comparisons of our X-ray diffraction with the thermal equation of state of *hcp*-Fe (ref. 32).

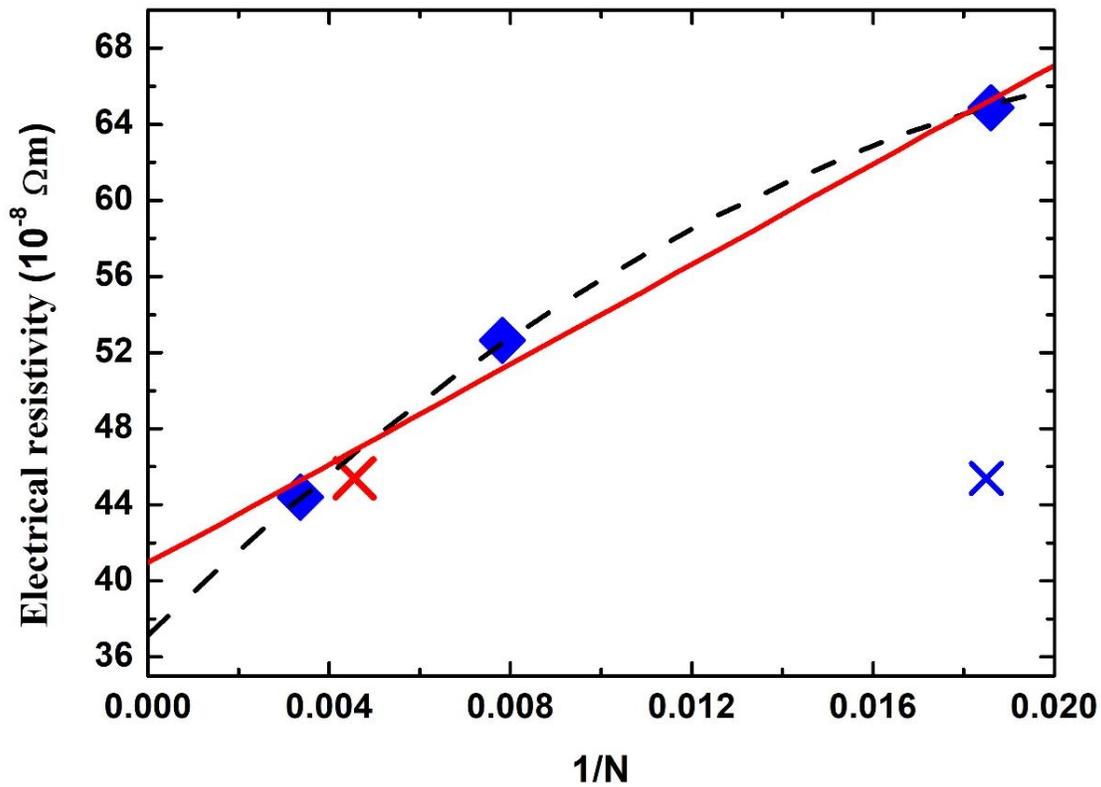

**Extended Data Figure 7 | Finite size extrapolation of resistivity computed exactly using the Kubo formalism within KKR-DMFT for snapshots from FPMD at 11.00 g/cm³ at 2,000 K in *hcp* iron.** The blue diamonds are for the Γ point in the FPMD. A 4×4×2 k-point set was used for the transport computations. The blue X is for 4 k-points (2×2×2) in a 54 atom cell. If this cell size is multiplied by 4 (red X) it falls on the relationship for the Γ point results, showing consistency, and that k-point sampling is the primary issue with the need for large cell sizes using only Γ. The red line gives 41×10⁻⁸ Ωm at infinite N, and the dashed black parabola gives 37×10⁻⁸ Ωm, whereas the actual results at 288 atoms is 44×10⁻⁸ Ωm, which gives some idea of the finite size uncertainty.

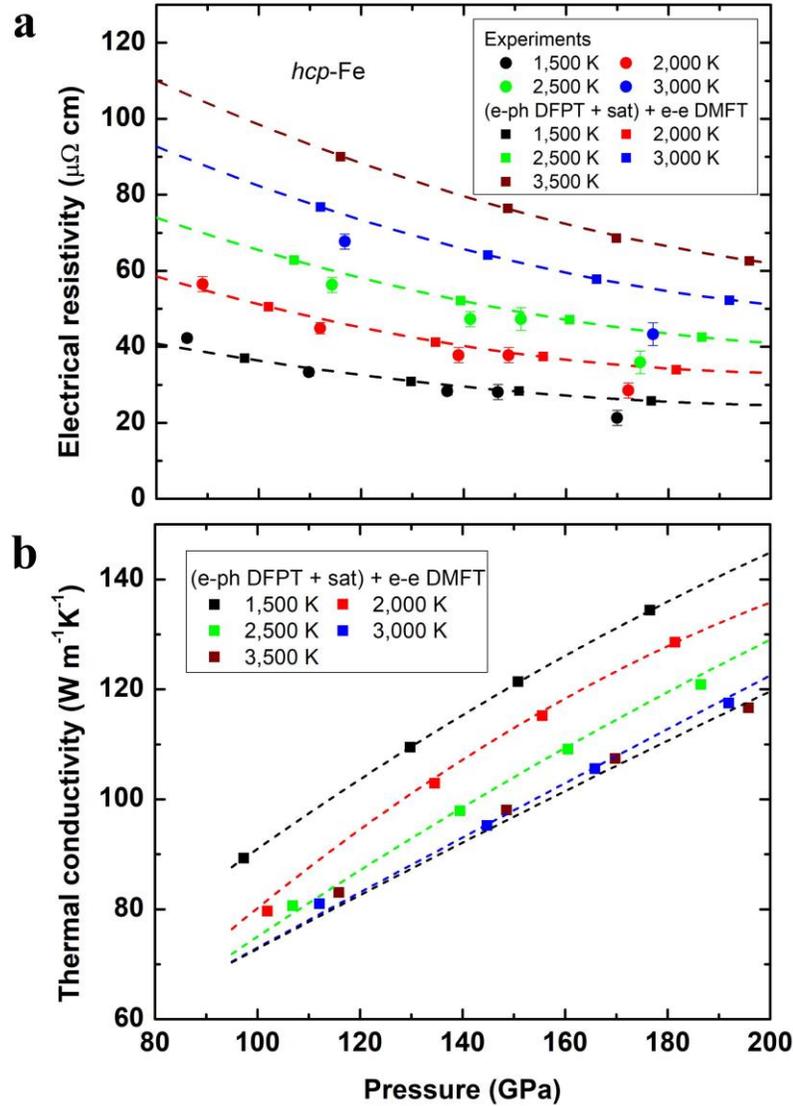

**Extended Data Figure 8 | Electrical resistivity and thermal conductivity versus pressure in *hcp*-Fe at high temperatures.** The measured and computed electrical resistivity (**a**) and calculated thermal conductivity (**b**) as functions of pressure and temperature (from 1,500 K to 3,500 K). The calculated resistivity and thermal conductivity of *hcp*-Fe are from Ref. 11 at constant volumes of 57.9, 54.9, 53.3, and 51.6 Bohr per atom, respectively. The resistivity of *hcp*-Fe decreases with increasing pressure and increases with increasing temperature. The dashed lines through the calculated points in the resistivity and thermal conductivity with increasing pressures are guides to the eye.

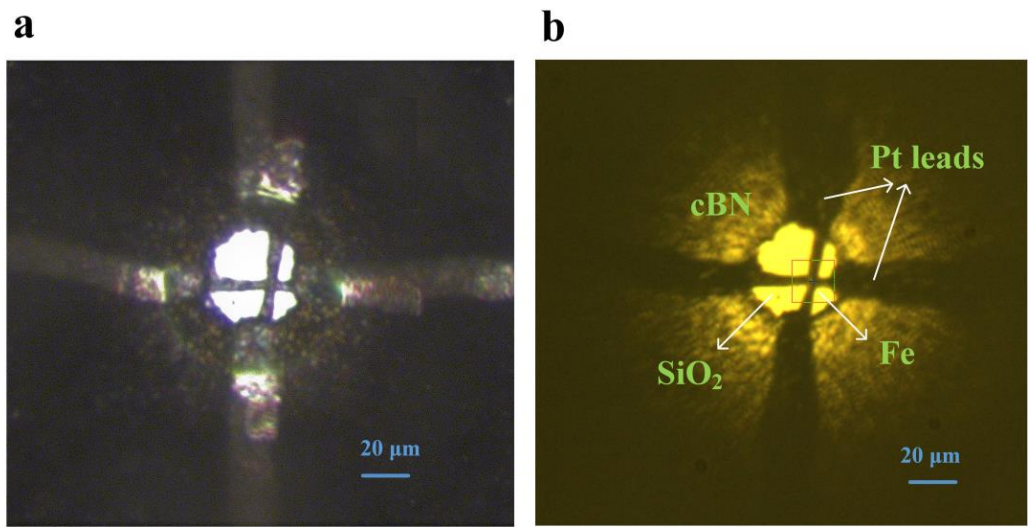

**Extended Data Figure 9 | Photographs of Fe sample loaded in a DAC at ~28 GPa (a) and ~142 GPa (b) and room temperature.** The images show that the sample was compressed uniformly in the chamber up to 142 GPa. *In situ* X-ray diffractions were also collected as shown in Extended Data Fig. 1a.

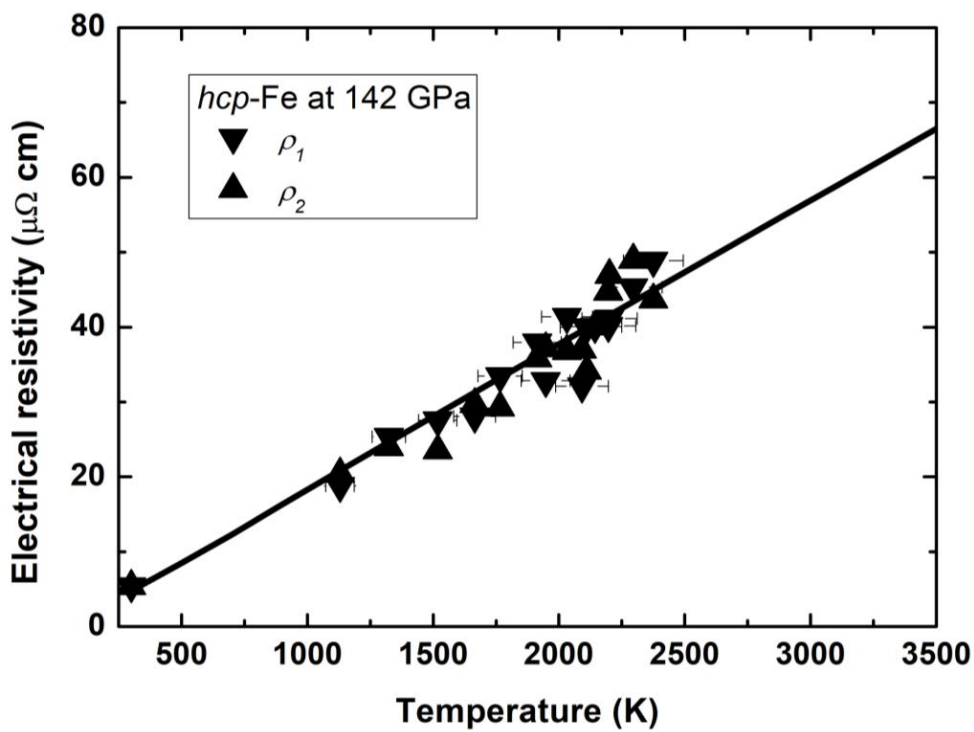

**Extended Data Figure 10 | Two sets of measured resistivity in *hcp*-Fe with increasing temperature at ~142 GPa.** Temperature response in resistivity $\rho_1$ and $\rho_2$ in *hcp*-Fe were obtained

when the direct current went through AB and CD path in Fig. 1c, respectively. The black solid line is the fitted line by the Block-Grüneisen formula for the average resistivity as shown in Extended Data Fig. 6c.

**Extended Data Table 1 | Fitted parameters for the Block-Grüneisen formula at high *P-T*.**

| Pressure (GPa) | Debye temperature $\theta_D$ (K) | $D(V)$ | $n$ |
|---|---|---|---|
| 74* | 608 | 17.3(0.2) | 3.658(0.325) |
| 82 | 620 | 17.7(0.3) | 3.539(0.221) |
| 105 | 655 | 14.8(0.1) | 1.565(0.050) |
| 133 | 692 | 13.1(0.3) | 1.001(0.021) |
| 142 | 703 | 13.4(0.2) | 1.005(0.025) |
| 165 | 730 | 10.6(0.2) | 0.806(0.025) |

\* Test run; $\theta_D$ was calculated from the equation of state of *hcp*-Fe[55].

**Extended Data Table 2. | Parameters used for modeling the energy balance of the present core.**

| Parameters | Variable | Value used |
|---|---|---|
| Core mass | $M_C$ | $1.9 \times 10^{24}$ kg |
| Present-day temperature at CMB | $T_{CMB}$ | ~3,900 K[45] |
| Present-day change rate of the core temperature | $\dot{T}_C$ | ~(-50) K/latest 0.5 Gyr* |
| Latent heat of Fe alloy at ICB | $L$ | ~500–600 kJ/kg[58] |
| Core specific heat capacity | $C_C$ | ~800 J/K kg[8,20] |
| Mass change of the inner core | $\dot{M}_{IC}$ | ~$2.1 \times 10^{14}$ kg/year |
| Core growth rate | - | ~0.9 mm/year (This study) |
| Density at the top of the inner core | - | $12.8 \times 10^3$ kg/m³ |

*Taken from the modeled evolution of core temperature in the study of Driscoll and Bercovici (2014).

## SI References